\documentclass[twocolumn,showpacs,preprintnumbers,amsmath,amssymb]{revtex4}
\usepackage{graphicx}% Include figure files
\usepackage{dcolumn}% Align table columns on decimal point
\usepackage{bm}% bold math

\begin{document}

%\preprint{APS/123-QED}

\title{Division of the Energy and of the Momentum of Electromagnetic Waves in Linear Media into Electromagnetic and Material Parts}

\author{Pablo L. Saldanha} \email{p.saldanha1@physics.ox.ac.uk}
 %\altaffiliation[Also at ]{Physics Department, XYZ University.}%Lines break automatically or can be forced with \\
\affiliation{Departamento de F\'isica, Universidade Federal de Minas
Gerais, Caixa Postal 702, 30123-970, Belo Horizonte, Minas Gerais,
Brazil. }
\affiliation{Department of Physics, University of Oxford, Clarendon Laboratory, Oxford, OX1 3PU, United Kingdom}

%\author{Charlie Author}
% \homepage{http://www.Second.institution.edu/~Charlie.Author}
%\affiliation{
%Second institution and/or address\\
%This line break forced% with \\
%}%

%\date{\today}% It is always \today, today,
             %  but any date may be explicitly specified

\begin{abstract} 
We defend a natural division of the energy density, energy flux and momentum density of electromagnetic waves in linear media in electromagnetic and material parts. In this division, the electromagnetic part of these quantities have the same form as in vacuum when written in terms of the macroscopic electric and magnetic fields, the material momentum is calculated directly from the Lorentz force that acts on the charges of the medium, the material energy is the sum of the kinetic and potential energies of the charges of the medium and the material energy flux results from the interaction of the electric field with the magnetized medium. We present reasonable models for linear dispersive non-absorptive  dielectric and magnetic media that agree with this division. We also argue that the electromagnetic momentum of our division can be associated with the electromagnetic relativistic momentum, inspired on the recent work of Barnett [Phys. Rev. Lett. \textbf{{104}}, 070401 (2010)] that showed that the Abraham momentum is associated with the kinetic momentum and the Minkowski momentum is associated with the canonical momentum. 
\end{abstract}

% PACS, the Physics and Astronomy
                             % Classification Scheme.
%\keywords{Suggested keywords}%Use showkeys class option if keyword
                              %display desired
\maketitle

\section{Introduction} 

The discussion of the correct form of the energy-momentum tensor of electromagnetic waves in material media has been the subject of a hot debate for more than one century \cite{pfeifer07}. The two most famous proposals were made by Minkowski and Abraham on the beginning of last century \cite{brevik79} and have distinct predictions for the momentum of light in a medium. The Abraham formulation predicts that an electromagnetic wave with total momentum $\mathbf{p}$ in vacuum decreases its momentum to $\mathbf{p}/n$ after entering a dielectric medium with refraction index $n$, while the Minkowski formulation predicts that the momentum of the wave is increased to $n\mathbf{p}$ after entering the medium. Several experiments on this subject were made in the past 60 years \cite{jones51,jones78,ashkin73,walker75a,walker75b,walker77,gibson80,campbell05,she08}, many of them with the purpose of solving the debate, but the fact is that no decision can be made on experimental grounds. As it was pointed out by Penfield and Haus \cite{penfield}, the Abraham, Minkowski and other electromagnetic energy-momentum tensors are accompanied by distinct material energy-momentum tensors. And when the properties of the material energy-momentum tensors are carefully taken into account, both formulations predict the same experimental results. A recent review of this debate can be seen in Ref. \cite{pfeifer07}. The eventual conclusion is that there are many compatible ways of dividing the total energy-momentum tensor of the system into an electromagnetic tensor and a material one.

Nevertheless, we can ask whether there is, among all possible ways of dividing the momentum and the energy of electromagnetic waves in material media into electromagnetic and material parts, a division that could be considered as more natural. Recently, many attempts have been made to calculate the momentum transfer from  electromagnetic waves to material media directly from the Lorentz force law \cite{gordon73,loudon02,loudon03,mansuripur04,scalora06,barnett06,mansuripur07b,hinds09,mansuripur09d,mansuripur10,saldanha10-mom,baxter10}. We believe that this is the most natural way of calculating the material momentum. In Ref. \cite{baxter10} it is presented a review of some treatments based on Lorentz force calculations for experiments on the topic, showing that consistent results are obtained.  We also believe that the most natural form for the material energy density is the sum of the potential and kinetic energy densities of the charges of the medium, which can be modified by the presence of an electromagnetic wave. The goal of this paper is to show, using reasonable models for material media, that if the material parts of the momentum density and of the energy density of electromagnetic waves in linear media are to be calculated in this way, the electromagnetic parts of the energy density $u_{\mathrm{e.m.}}$, of the energy flux $\mathbf{S}_{\mathrm{e.m.}}$ and of the momentum density $\mathbf{p}_{\mathrm{e.m.}}$ must be
\begin{eqnarray}\label{energy em}	&&u_{\mathrm{e.m.}}=\frac{\varepsilon_0}{2}|\mathbf{E}|^2+\frac{1}{2\mu_0}|\mathbf{B}|^2\;,\;\mathbf{S}_{\mathrm{e.m.}}=\frac{1}{\mu_0}\mathbf{E}\times\mathbf{B}\;,\\\label{mom em}
	&&\mathbf{p}_{\mathrm{e.m.}}=\varepsilon_0\mathbf{E}\times\mathbf{B}\;,
\end{eqnarray}
where $\mathbf{E}$ represents the macroscopic electric field, $\mathbf{B}$ the macroscopic magnetic field, $\varepsilon_0$ the permittivity of free space and $\mu_0$ the permeability of free space. These forms are identical to the expressions in vacuum, but written in terms of the macroscopic fields. The form for the electromagnetic momentum density above had appeared before, for instance in the book of Livens \cite{livens}. However, the discussion of the energy of electromagnetic waves in linear media is practically absent from the literature.

In Sec. \ref{discussion} our proposal for the division is more carefully described and compared with previous treatments. In Secs. \ref{sec:clas-diel}, \ref{sec:quant-diel} and \ref{sec:quant-mag} different models for dispersive non-absorptive dielectric and magnetic media are shown to be consistent with the proposed material energy density. In Sec. \ref{sec:flux} we reproduce the model introduced by Penfield and Haus \cite{penfield7-4} for the material energy flux and material momentum of an object with magnetic dipole moment in an applied electric field, showing the consistency of the division. This momentum of a magnetic dipole in an electric field is frequently called ``hidden momentum'' \cite{schockley67,vaidman90} and is a relativistic effect. In recent papers, Barnett \textit{et al.} \cite{hinds09,barnett10a,barnett10b} associated the Minkowski momentum with the electromagnetic canonical momentum and the Abraham momentum with the electromagnetic kinetic momentum. Based on the model of Penfield and Haus, we suggest that the momentum density $\mathbf{p}_{\mathrm{e.m.}}$ of Eq. (\ref{mom em}) should be associated with the electromagnetic relativistic momentum. Finally, in Sec. \ref{sec:conclusions} we present our concluding remarks.

\section{Discussion of the division}\label{discussion}

The microscopic Lorentz force density that acts on a density charge $\rho'$ moving with velocity $\mathbf{v}$, forming a current density $\mathbf{j}'=\rho' \mathbf{v}$, is
\begin{equation}
	\mathbf{f}'=\mathbf{e}\rho'+\mathbf{j}'\times \mathbf{b}\;,
\end{equation}
where $\mathbf{e}$ and $\mathbf{b}$ are the microscopic electric and magnetic fields. If the separation between the charges in the medium is much smaller than the wavelength of the incident electromagnetic wave, we can associate averaged electric and magnetic fields $\mathbf{E}$ and $\mathbf{B}$ to volumes containing a huge number of particles, but still with dimensions much smaller than the radiation wavelength. The microscopic electric and magnetic fields can be written as $\mathbf{e}=\mathbf{E}+{\mbox{\boldmath{$\delta$}}}_e$, $\mathbf{b}=\mathbf{B}+{\mbox{\boldmath{$\delta$}}}_b$, where ${\mbox{\boldmath{$\delta$}}}_e$ and ${\mbox{\boldmath{$\delta$}}}_b$ depend on the microscopic distribution of charges and average to zero. So the averaged force density in this volume can be written as 
\begin{equation}\label{lorentz}
	\mathbf{f}=\mathbf{E}\rho+\mathbf{J}\times \mathbf{B}\;,
\end{equation}
where $\rho$ represents the averaged charge density and $\mathbf{J}$ the averaged current density in the volume. In a linear medium without free charges, we can write $\mathbf{J}=\partial \mathbf{P}/\partial t+\nabla\times \mathbf{M}$ and $\rho=-\nabla\cdot \mathbf{P}$, where $\mathbf{P}$ and $\mathbf{M}$ are the polarization and magnetization of the medium \cite{jackson}. The electric and magnetic macroscopic fields are the fields that appear on the Lorentz force equation above, so these are the fields that can transfer momentum and energy to material macroscopic bodies. From our point of view, these should be the fields that transport electromagnetic energy and momentum. For this reason, these are the fields that appear in Eqs. (\ref{energy em}) and (\ref{mom em}) for the electromagnetic energy density, energy flux and momentum density. The Abraham and Minkowski momentum densities, on the other hand, are $\mathbf{p}_{\mathrm{Abr}}=\mathbf{E}\times \mathbf{H}/c^2$ and $\mathbf{p}_\mathrm{Min}=\mathbf{D}\times \mathbf{B}$, respectively, where $\mathbf{H}\equiv \mathbf{B}/\mu_0-\mathbf{M}$, $\mathbf{D}\equiv \varepsilon_0\mathbf{E}+\mathbf{P}$ and $c=1/\sqrt{\mu_0\varepsilon_0}$ is the speed of light in vacuum \cite{pfeifer07,brevik79}.

Some of the previously cited papers that calculate the material momentum by the Lorentz force consider that the electromagnetic momentum density should have the Abraham value $\mathbf{E}\times \mathbf{H}/c^2$ \cite{gordon73,scalora06,mansuripur09d,mansuripur10}. Our proposal for the electromagnetic momentum density of Eq. (\ref{mom em}) agrees with the Abraham form in a non-magnetic medium. As it was discussed by Mansuripur \cite{mansuripur09d,mansuripur10}, to have momentum conservation considering the Abraham form for the electromagnetic momentum density in magnetic media, a modification must be done in the Lorentz force law. Mansuripur opts for this modification in the Lorentz force to avoid the concept of ``hidden momentum'', that we will discuss in Sec. \ref{sec:flux}. In a recent paper \cite{saldanha10-mom}, we showed that the form of Eq. (\ref{mom em}) for the electromagnetic momentum density and the form of Eq. (\ref{lorentz}) for the Lorentz force law are compatible with momentum conservation in a series of examples \footnote{Actually we used a different form for the force density in Ref. \cite{saldanha10-mom}. But in Ref. \cite{barnett06} it is shown that the force density we used is equivalent to the one of Eq. (\ref{lorentz}).}. We also showed compatibility with the Balazs \textit{gedanken} experiment \cite{balazs53}, showing compatibility with the theory of relativity when taking into account the ``hidden momentum''. Because in our formulation there is no need for a modification in the Lorentz force law, we believe that our treatment is more natural.

Now let us discuss the energy of electromagnetic waves in linear media. In a closed system, the local conservation of energy can be written as
\begin{equation}
	\frac{\partial u_{\mathrm{tot}}}{\partial t}=-\nabla\cdot \mathbf{S}_{\mathrm{tot}}\;,
\end{equation}
where $u_{\mathrm{tot}}$ and $\mathbf{S}_{\mathrm{tot}}$ represent the total energy density and total energy flux of the system. For a system composed by an approximately monochromatic electromagnetic wave with frequency $\omega_c$ in a linear dispersive non-absorptive dielectric and magnetic medium, we can write \cite{jackson}
\begin{eqnarray}\nonumber
	\langle u_{\mathrm{tot}}\rangle=&&\left[1+\frac{d(\omega\chi_\mathrm{e})}{d\omega}\right]_{\omega=\omega_c}
    \frac{\varepsilon_0\langle|\mathbf{E}|^2\rangle}{2}+\\&&+\left[1+\frac{d(\omega\chi_\mathrm{m})}{d\omega}\right]_{\omega=\omega_c}
    \frac{\mu_0\langle|\mathbf{H}|^2\rangle}{2}\;,\\
    \mathbf{S}_{\mathrm{tot}} =&&\mathbf{E}\times \mathbf{H}\;,
\end{eqnarray}
where $\langle A\rangle$ represents the average of $A$ over one period of oscillation of the field, $\chi_\mathrm{e}(\omega)$ is the electric susceptibility and $\chi_\mathrm{m}(\omega)$ is the magnetic susceptibility of the medium. So, if the electromagnetic parts of the energy density and of the energy flux of electromagnetic waves in linear media are given by Eq. (\ref{energy em}), the material counterparts must be given by
\begin{eqnarray}\label{dens en mat}\nonumber
    \langle u_{\mathrm{mat}}\rangle=&&  \left[\chi_\mathrm{e}+\omega\frac{d\chi_\mathrm{e}}{d\omega}\right]_{\omega=\omega_c}
    \frac{\varepsilon_0\langle|\mathbf{E}|^2\rangle}{2}+\\&&+\left[-\alpha_\mathrm{m}+\omega\frac{d\alpha_\mathrm{m}}{d\omega}\right]_{\omega=\omega_c}
    \frac{\langle|\mathbf{B}|^2\rangle}{2\mu_0}\;,\\\label{flux mat}
    \mathbf{S}_{\mathrm{mat}} =&&-\mathbf{E}\times \mathbf{M}\;,
\end{eqnarray}
with $\alpha_\mathrm{m}\equiv \chi_\mathrm{m}/(1+\chi_\mathrm{m})$, so that $\langle u_{\mathrm{e.m.}} +u_{\mathrm{mat}}\rangle =\langle u_{\mathrm{tot}}\rangle$ and $\mathbf{S}_{\mathrm{e.m.}} +\mathbf{S}_{\mathrm{mat}} = \mathbf{S}_{\mathrm{tot}}$. In the next three sections we show that commonly used models for dielectric and magnetic linear media acquire the energy density of Eq. (\ref{dens en mat}) with the presence of an electromagnetic wave, and in Sec. \ref{sec:flux} we reproduce the model of Penfield and Haus \cite{penfield7-4} that justifies the material energy flux of Eq. (\ref{flux mat}).

\section{Classical model for a linear dielectric medium} \label{sec:clas-diel}

The first model that we present is the Drude-Lorentz model for dielectric media. A similar treatment was previously made by Haus and Kogelnik \cite{haus75}. In this model, the atomic nuclei are fixed and the electrons are bound by harmonic potentials $U_i(r_i)=m\omega_i^2r_i^2/2$, where $m$ represents the electron mass, $r_i$ the distance of each electron to its equilibrium position and $\omega_i$ the natural frequency of the harmonic oscillator. Considering a non-absorptive medium, the equation of motion of the $\mathbf{\hat{x}}$ component of each electron can be written as 
\begin{equation}
m\frac{d^2x_i}{dt^2}=qE\cos(\omega t)-\frac{\partial U_i}{\partial x_i}\;,
\end{equation} where $q$ represents the electron charge and $\mathbf{E}={E}\cos(\omega t)\mathbf{\hat{x}}$ is the incident electric field. The solutions of the above differential equation are
\begin{equation}\label{x}
x_i(t)=\frac{q}{m(\omega_i^2-\omega^2)}{E}\cos(\omega t)\;.
\end{equation}
Because we are disregarding the absorption of radiation, we must consider that $\omega$ is far from  $\omega_i$. 

If we call $N_i$ the density of electrons subjected to the potential $U_i$, the electric susceptibility of the medium can be written as
\begin{equation}\label{chi_clas}
    \chi_e(\omega)=\frac{|\mathbf{P}|}{\varepsilon_0
    |\mathbf{E}|}=\frac{\sum_iN_iqx_i
    }{\varepsilon_0E\cos(\omega t)}=\sum_i\frac{N_iq^2}{\varepsilon_0m(\omega_i^2-\omega^2)}.
\end{equation} 

The material energy density corresponds tho the sum of the kinetic $(m\dot{x}_i^2/2)$ and potential $(m\omega_i^2x_i^2/2)$ energies of the electrons multiplied by their densities $N_i$. Using Eq. (\ref{x}), we have
 \begin{equation}\label{dens en
mat 1}
    u_\mathrm{mat}=
    \sum_i\frac{N_iq^2(\omega_i^2+\omega^2)}{m(\omega_i^2-\omega^2)^2}
    \frac{|\mathbf{E}|^2}{2}.
\end{equation}

Using Eq. (\ref{chi_clas}), it is straightforward to see that Eq. (\ref{dens en mat 1}) can be written as the first term of Eq. (\ref{dens en mat}):
 \begin{equation}
    u_{\mathrm{mat}}= \left[\chi_\mathrm{e}+\omega\frac{d\chi_\mathrm{e}}{d\omega}\right]
    \frac{\varepsilon_0|\mathbf{E}|^2}{2}.
\end{equation}

\section{Quantum model for a linear dielectric medium}\label{sec:quant-diel}

Now let us consider a quantum model for a linear non-absorptive dielectric medium. The medium is composed by molecules with fixed nuclei and $Z$ electrons with charge $q$ and mass $m$ around. The Hamiltonian that couples these nuclei and electrons may be very complex, but our results do not depend on its specific form. The eigenstates of the molecular Hamiltonian will be denoted $|\Phi_n\rangle$, and the respective eigenvalues $\hbar\omega_n$. An arbitrary pure state of the molecular system can be written as
\begin{equation}\label{estado atomo}
|\Phi(t)\rangle=\sum_nc_n(t)|\Phi_n\rangle\mathrm{e}^{-{i\omega_nt}}\;,
\end{equation}
with $\sum_n|c_n(t)|^2=1$.

An electric field $\mathbf{E}=E\cos(\omega t)\mathbf{\hat{x}}$ acts as a perturbation on the system. Our treatment is based on Loudon's book \cite{loudon}. The total Hamiltonian can be written as
\begin{eqnarray}\label{Hamilt pert}
    &&\hat{H}=\hat{H}_0+\hat{H}'(t)\;,\;\mathrm{with}\;\;\hat{H}_0|\Phi_n\rangle=\hbar\omega_n|\Phi_n\rangle,\\
    &&\hat{H}'(t)=-E\cos(\omega
    t)Zq\hat{x}_\mathrm{cm},
\end{eqnarray} where $\hat{x}_\mathrm{cm}$ is the $\mathbf{\hat{x}}$ component of the center of mass position operator of the $Z$ electrons  and $Zq$ is the total charge of the electrons. 

From the Schr\"odinger equation with the Hamiltonian above applied on the state (\ref{estado atomo}), after taking the scalar product on both sides with $\mathrm{e}^{i{\omega_mt}}\langle \Phi_m|$, we obtain
\begin{equation}\label{eq dif cm}
    \dot{c}_m(t)=-\frac{i}{\hbar}\sum_n
    c_n(t)\langle\Phi_m|\hat{H}'|\Phi_n\rangle\mathrm{e}^{i{(\omega_m-\omega_n)t}}\;.
\end{equation}

Let us consider that the molecule is in its fundamental state
$|\Phi_0\rangle$ before the interaction with the field starts and that the field produces a small perturbation, such that
 $c_0\approx 1$ always. So we will consider only the term $n=0$ in the summation above to the calculus of $\dot{c}_m(t)$. The electric susceptibility refers to the situation where the interaction between the field and the molecules has achieved an equilibrium condition, so we can take an indefinite integral as the solution of Eq. (\ref{eq dif cm}): \begin{equation}\label{cm}
c_m(t)\simeq\frac{EZqX_{m0}}{2\hbar}\left[\frac{\mathrm{e}^{i(\omega_m+\omega)t}}{\omega_m+\omega}+
    \frac{\mathrm{e}^{i(\omega_m-\omega)t}}{\omega_m-\omega}\right],
\end{equation} with
$X_{m0}\equiv\langle\Phi_m|\hat{x}_{\mathrm{cm}}|\Phi_0\rangle$,
$\omega_0\equiv0$, $c_0\approx 1$. We used the definition
$X_{00}\equiv0$ to define the origin of the coordinates system.

 The oscillating dipole moment of the molecule can be written as
\begin{eqnarray}\label{dip}
  d_x(t)&=&\langle\Phi(t)|Zq\hat{x}_\mathrm{cm}|\Phi(t)\rangle\nonumber\\
  &\simeq&\sum_mc_m(t)ZqX_{m0}^*\mathrm{e}^{-i\omega_m
    t}+\mathrm{c.c.},
\end{eqnarray} where we disregarded terms  $c_mc_n$ with $m$ and $n$ both different from zero, $c_0$ was considered 1 and c.c. stands for the complex conjugate.

Using Eqs. (\ref{dip}) and (\ref{cm}), we can see that a medium composed by a density $N$ of these molecules has an electric susceptibility
 \begin{equation}\label{chi}
\chi_\mathrm{e}(\omega)=\frac{Nd_x}{\varepsilon_0|\mathbf{E}|}=\frac{NZ^2q^2}{\varepsilon_0\hbar}\sum_m|X_{m0}|^2\left[\frac{1}
{\omega_m+\omega}+\frac{1} {\omega_m-\omega}\right].
\end{equation} Because we are considering a non-absorptive medium, $\chi_e(\omega)$ is real. This approximation of non-absorptive media is valid only when $\omega$ is far from all $\omega_m$ for which $X_{m0}\neq0$.

The material energy density is the expectation value of the non-perturbed Hamiltonian $\hat{H}_0$ multiplied by the density of molecules, because the Hamiltonian  $\hat{H}_0$ contains the kinetic and potential energy of the electrons:
\begin{equation}\label{dens en atomos}
    u_{\mathrm{mat}}=N\langle\Phi|\hat{H}_0|\Phi\rangle=N\sum_m|c_m|^2
    \hbar\omega_m.\end{equation}

Using Eq. \ref{cm}, we have \begin{equation}
|c_m(t)|^2\simeq\frac{Z^2q^2E^2|X_{m0}|^2}{\hbar^2(\omega_m^2-\omega^2)^2}\left[\omega_m^2\cos^2(\omega
t)+\omega^2\sin^2(\omega t)\right]. \end{equation}

Taking the temporal average over one period of oscillation of the field we have
\begin{equation}\label{dens en atom final}
   \langle u_{\mathrm{mat}}\rangle\simeq\frac{NZ^2q^2}{\hbar}\sum_m\frac{\omega_m|X_{m0}|^2(\omega_m^2+\omega^2)}
{(\omega_m^2-\omega^2)^2}\langle|\mathbf{E}|^2\rangle.
\end{equation}

Using Eq. (\ref{chi}), it is straightforward to show that the energy density can be written as the first term of Eq.  (\ref{dens en mat}):
 \begin{equation}\label{dens en mat quant}
    \langle u_{\mathrm{mat}}\rangle= \left[\chi_\mathrm{e}+\omega\frac{d\chi_\mathrm{e}}{d\omega}\right]
    \frac{\varepsilon_0\langle|\mathbf{E}|^2\rangle}{2}.
\end{equation}

If we have different kinds of molecules composing the medium, each one identified by a symbol $(j)$, the electric susceptibility and the energy density of the medium will be \begin{equation}
    \chi_\mathrm{e}^{Tot}=\sum\chi_\mathrm{e}^{(j)},\;\;\;\;\langle u_{\mathrm{mat}}^{Tot}\rangle=\sum
    \langle u_{\mathrm{mat}}^{(j)}\rangle,
\end{equation} and Eq. (\ref{dens en mat quant}) remains valid.

\section{Quantum model for a linear magnetic medium}\label{sec:quant-mag}

The quantum model that we  use for a magnetic linear medium is based on Van Vleck's book \cite{vanvleck} and considers the response of electrons under the action of a central potential to an applied magnetic field. We use cylindrical coordinates $(z,s,\phi)$ in this section. To simplify the treatment, we will consider that the magnetic field is generated by a vector potential $\mathbf{A}=Bs\mathbf{\hat{\mathbf{\phi}}}/2$, such that
$\mathbf{B}=\nabla \times A=B\mathbf{\hat{z}}$, instead of considering the vector potential of an electromagnetic wave. The Lagrangean of an electron of mass $m$ and charge $q$ under the action of this magnetic field and an atomic central potential $\Phi(z,s)$ is
 \begin{eqnarray}\label{lagrange magnet}\nonumber
    \mathcal{L}&=&\frac{1}{2}m|\dot{\mathbf{r}}|^2+q\mathbf{A}\cdot\dot{\mathbf{r}}-q\Phi(z,s)\\    
    &=&\frac{m\dot{z}^2}{2}+\frac{m\dot{s}^2}{2}+\frac{ms^2\dot{\phi}^2}{2}+\frac{qs^2B\dot{\phi}}{2}-q\Phi(z,s).
\end{eqnarray}

The Lagrangean does not depend on the variable $\phi$, so the momentum conjugated to this coordinate is a constant of the movement of the system: \begin{equation}
    p_\phi=\frac{\partial \mathcal{L}}{\partial
    \dot{\phi}}=ms^2\dot{\phi}+\frac{qs^2B}{2}=\;\mathrm{constant}.
\end{equation} So the $\mathbf{\hat{z}}$ component of the angular momentum of the electron, $l_z=ms^2\dot{\phi}$, is a function of the applied magnetic field: \begin{equation}
    l_z(B)=l_z(0)-\frac{qs^2B}{2},
\end{equation} where $l_z(0)$ represents the original angular momentum of the electron, when the magnetic field is zero.

The ratio between the dipole moment and the orbital angular momentum of an electron is $q/(2m)$ \cite{jackson}, so the variation of the electron dipole moment is $\Delta\mu_z=\mu_z(B)-\mu_z(0)=-{q^2}\langle s^2\rangle B/{4m}$,
 with $\langle s^2\rangle\equiv\langle\Psi|
\hat{s}^2|\Psi\rangle$, $|\Psi\rangle$ being the unperturbed quantum state of the electron. Considering that the medium has atoms with random orientations, on taking the average of the orientations we can consider, for each quantum sate of the electrons, $\sum_j\langle l_z(0)\rangle_j=0$ and $\sum_j\langle
s^2\rangle_j=\sum_j 2/3\langle r^2\rangle_j$, where $r$ is the distance from the origin of the central potential. Considering a density $N_i$ for electrons in each quantum state $|\Psi_i\rangle$, the magnetization of the medium can be written as
\begin{equation}
    \mathbf{M}=\sum_i-\frac{N_iq^2\langle r^2\rangle_i}{6m}\mathbf{B}.
\end{equation} So the coefficient $\alpha_\mathrm{m}$, defined by the relation $\mathbf{M}=\alpha_\mathrm{m} \mathbf{B}/\mu_0=\chi_\mathrm{m} \mathbf{H}$, so that $\alpha_\mathrm{m}=\chi_\mathrm{m}/(1+\chi_\mathrm{m})$, is
\begin{equation}\label{alpham}
    \alpha_\mathrm{m}=\sum_i-\frac{\mu_0N_iq^2\langle r^2\rangle_i}{6m}.
\end{equation}

The variation of the electrons kinetic energy density can be written as 
\begin{eqnarray}
    u_\mathrm{mat}&=&\sum_iN_i\left\langle
    \frac{l_z(B)^2-l_z(0)^2}{2ms^2}\right\rangle_i\nonumber \\
    &=&\sum_iN_i\left\langle
    \frac{q^2r^2B^2}{12m}\right\rangle_i.
\end{eqnarray}  
Using Eq. (\ref{alpham}), we can see that this energy density can be written as the second term of Eq. (\ref{dens en mat}), \begin{equation}
    u_\mathrm{mat}=-\alpha_m\frac{B^2}{2\mu_0},
\end{equation} since in this model $\alpha_m$ does not depend on the frequency $\omega$.

\section{Classical model for the material energy flux and ``hidden momentum''}\label{sec:flux}

\begin{figure} \begin{center}
  % Requires \usepackage{graphicx}
  \includegraphics[width=7cm]{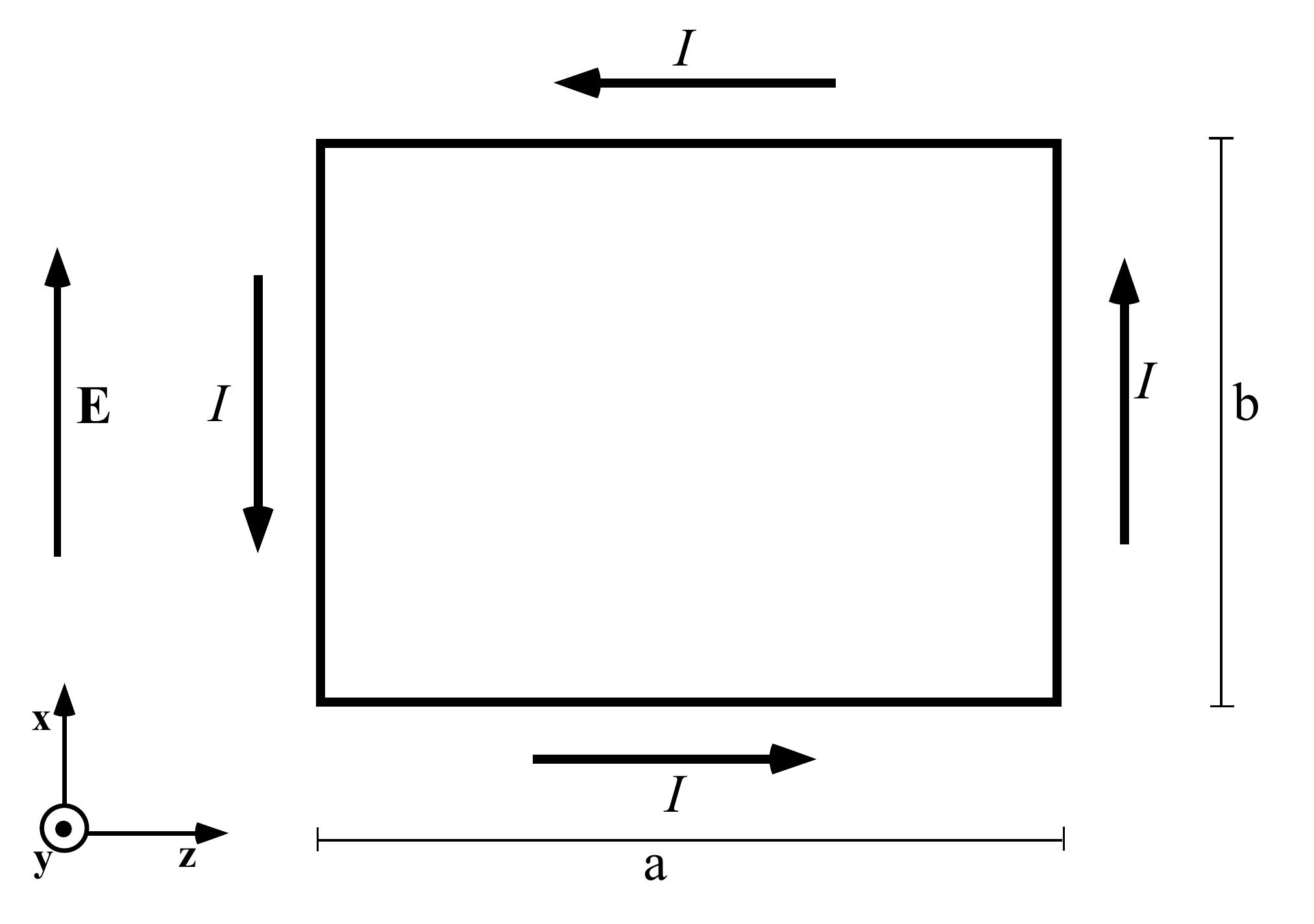}\\
  \caption{Magnetic dipole in an electric field has associated energy flux and linear momentum.}\label{fig-hidden}
\end{center}
 \end{figure}

Here we reproduce the model of Penfield and Haus \cite{penfield7-4} that shows that a magnetic dipole under the action of an electric field can have associated energy flux and momentum, even if the dipole is not moving. Consider a rectangle of sides $a$ and $b$ that conducts an electric current $I$ in a region of space with an electric field $\mathbf{E}$. The situation is depicted in Fig. \ref{fig-hidden}. The magnetic dipole moment of the object is $\mathbf{m}=abI\mathbf{\hat{y}}$. The electric field does a positive work on the positive charges that go up in the right side of the figure and a negative work on the positive charges that go down in the left side (if we consider that the moving charges are negative, the final conclusions will be the same). So a charge $q$ moving to the left has an extra energy $qEb$ with respect to the situation when it moves to the right. This fact generates an energy flux from right to left. If we have a density $N$ of these dipoles, the energy flux can be found multiplying the energy per unity time $IEb$ that goes to the left in each dipole by the length $a$ along which the energy flows and by the dipole density $N$. So we have
\begin{equation}\label{fluxo en mat modelo}
    \mathbf{S}_\mathrm{mat}=-IEbaN\mathbf{\hat{z}}=-\mathbf{E}\times \mathbf{M}\;,
\end{equation} where $\mathbf{M}$ is the magnetization of the medium. This result agrees with Eq. (\ref{flux mat}).

The simple model presented in this section is very distinct from an atomic system. However, since the magnetic response of an atom due to a magnetic field oscillating at optical frequencies is due to the orbital motion of the electrons, a similar phenomenon should occur in that case.

Now let us analyze the relativistic linear momentum of the object. The momentum of the charges that move to the left in Fig. \ref{fig-hidden} is higher because they have more energy. The ratio between the momentum and the energy of a relativistic particle is $\mathbf{v}/c^2$, where $\mathbf{v}$ is the particle velocity. Let us consider that the current $I$ is uniform along the circuit. So if we have a linear  charge density $\lambda_l$ with velocity  $v_l$ going to the left and a linear  charge density $\lambda_r$ with velocity $v_r$ going to the right, we have $I=\lambda_lv_l=\lambda_rv_r$. Calling $U$ the energy of a particle that goes to the right in Fig. \ref{fig-hidden} and $q$ the total moving charge of the circuit, the total linear momentum of the circuit is
\begin{eqnarray} \mathbf{P}&=&\left[\frac{\lambda_rv_raU}{qc^2}-\frac{\lambda_lv_la[U+qEb]}{qc^2}\right]\mathbf{\hat{z}}
=-\frac{IabE}{c^2}\mathbf{\hat{z}}\nonumber \\ 
&=&\frac{\mathbf{m}\times
    \mathbf{E}}{c^2}.
\end{eqnarray} So, if we have a magnetization $\mathbf{M}$ in a medium with an electric field $\mathbf{E}$, there will be a ``hidden momentum'' density $\mathbf{p}_{\mathrm{hid}}=-\mathbf{E}\times\mathbf{M}/c^2$. It is worth mentioning that the ``hidden momentum'' is a relativistic effect. If we had considered that the momentum of a particle is $m\mathbf{v}$, there would be no ``hidden momentum''.

In recent works, Barnett \textit{et al.} \cite{hinds09,barnett10a,barnett10b} associated the Abraham momentum with the electromagnetic kinetic momentum and the Minkowski momentum with the electromagnetic canonical momentum. Considering the ``hidden momentum'', we see that the relativistic momentum of an atom with magnetic moment $\mathbf{m}$ can be written as $\mathbf{P}_\mathrm{rel}=\mathbf{P}_\mathrm{kin}-\mathbf{E}\times \mathbf{m}/c^2$, where $\mathbf{P}_\mathrm{kin}=m\mathbf{v}$, $m$ being the mass and $\mathbf{v}$ the velocity of the atom. If we have a bunch of these atoms forming a medium, the difference of the relativistic and the kinetic momenta of all atoms is equal to the difference of the total momentum of the field when we consider the Abraham momentum $\mathbf{P}_\mathrm{Abr}$, and the total momentum $\mathbf{P}_\mathrm{e.m.}$ when we consider the momentum density of Eq. (\ref{mom em}). So we can write
\begin{equation}	\mathbf{P}_\mathrm{e.m.}+\mathbf{P}_\mathrm{rel}^{\mathrm{med}}=\mathbf{P}_\mathrm{Abr}+\mathbf{P}_\mathrm{kin}^{\mathrm{med}}=\mathbf{P}_\mathrm{Min}+\mathbf{P}_\mathrm{can}^{\mathrm{med}}.
\end{equation} The last equality is derived in Ref. \cite{hinds09}. The total kinetic and canonical momenta are always conserved, so we conclude that, if we associate $\mathbf{P}_\mathrm{e.m.}$ with the relativistic momentum of the field, the total relativistic momentum will be always conserved.

\section{Conclusions}\label{sec:conclusions}

In this paper we defended a particular division of the energy and of the momentum of electromagnetic waves in linear media in electromagnetic and material parts as being the most natural. The electromagnetic parts of the energy density, energy flux and momentum density are given by Eqs. (\ref{energy em}) and (\ref{mom em}), having the same form than in vacuum when written in terms of the macroscopic electric and magnetic fields. The material momentum is calculated directly from the Lorentz force law, the material energy corresponds to the kinetic and potential energies of the charges of the medium and the material energy flux is the result of the work done by the electric field on the charges that compose an object with magnetic dipole moment. 

In a previous work \cite{saldanha10-mom}, we showed momentum conservation in various circumstances considering the electromagnetic momentum density of Eq. (\ref{mom em}) and calculating the material momentum with the Lorentz force law. Here we showed, using reasonable models for linear media, that we have energy conservation considering the energy density of Eq. (\ref{energy em}) and calculating the material energy density as the sum of the kinetic and potential energy densities of the charges of the medium. To justify the material energy flux, we used a model introduced by Penfield and Haus \cite{penfield7-4}. We also argued that the momentum density of Eq. (\ref{mom em}) can be associated with the relativistic electromagnetic momentum. It is worth to mention that, although we used models for linear non-absorptive media, we believe that the proposed division can be extended to any kind of media. For example, on an absorbing medium the absorbed energy is usually converted into heat or chemical reactions, that can also be associated with the kinetic and potential energies of the particles of the medium.

\begin{acknowledgments}

The author acknowledges Prof. C. H. Monken and Prof. Masud Mansuripur for very useful discussions. This work was supported by the Brazilian agencies CNPq and CAPES.

\end{acknowledgments}

%\bibliography{refer}% Produces the bibliography via BibTeX.

\end{document}